\documentclass{llncs}
\usepackage[utf8]{inputenc}
\usepackage{amssymb,amsmath,latexsym}
\usepackage{url}
\usepackage[pdftex]{graphicx}

\usepackage[plainpages=false, pdfpagelabels]{hyperref}
\usepackage[all,color,line]{xy}

\CompileMatrices

\usepackage{color}

\newcommand{\X}{\mathsf{X}}     
\newcommand{\U}{{\,\uU\,}}      
\newcommand{\uU}{\mathsf{U}}    
\newcommand{\rR}{\mathsf{R}}    
\newcommand{\R}{{\,\rR\,}}      
\newcommand{\F}{\mathsf{F}}     
\newcommand{\G}{\mathsf{G}}     

\newcommand{\true}{\textrm{{\it tt}}}

\newcommand{\AP}{\mathit{A\hskip-0.1ex P}}

\newcommand{\mA}{\mathcal{A}}
\newcommand{\mB}{\mathcal{B}}
\newcommand{\mG}{\mathcal{G}}
\newcommand{\mF}{\mathcal{F}}
\newcommand{\mT}{\mathcal{T}} 

\begin{document}

\frontmatter


\title{LTL to B\"{u}chi Automata Translation:\\ Fast and More Deterministic%
  \thanks{The authors are supported by The Czech Science Foundation, grants
    102/09/H042 (Babiak), 201/09/1389 (K\v{r}et\'{\i}nsk\'{y}),
    P202/10/1469 (\v{R}eh\'{a}k, Strej\v{c}ek), P202/12/G061
    (K\v{r}et\'{\i}nsk\'{y}, \v{R}eh\'{a}k, Strej\v{c}ek), and
  P202/12/P612 (\v{R}eh\'{a}k).}}
\author{Tom\'{a}\v{s} Babiak \and Mojm\'{i}r K\v{r}et\'{i}nsk\'{y} \and
  Vojt\v{e}ch~{\v{R}}eh\'{a}k \and Jan Strej\v{c}ek}
    
\institute{Faculty of Informatics, Masaryk University\\
  Botanick\'{a} 68a, 60200 Brno, Czech Republic \\
  \email{\{xbabiak,\,kretinsky,\,rehak,\,strejcek\}@fi.muni.cz}}


\maketitle


\begin{abstract}
  We introduce improvements in the algorithm by Gastin and Oddoux
  translating LTL formulae into B\"uchi automata via very weak alternating
  co-B\"uchi automata and generalized B\"uchi automata. Several improvements
  are based on specific properties of any formula where each branch of its
  syntax tree contains at least one \emph{eventually} operator and at least
  one \emph{always} operator. These changes usually result in faster
  translations and smaller automata. Other improvements reduce
  non-determinism in the produced automata. In fact, we modified all the
  steps of the original algorithm and its implementation known as
  LTL2BA. Experimental results show that our modifications are real
  improvements. Their implementations within an LTL2BA translation made
  LTL2BA very competitive with the current version of SPOT, sometimes
  outperforming it substantially.

  This is a full version of~\cite{tacas12} published at TACAS 2012.
\end{abstract}	


\section{Introduction}


A translation of LTL formulae into equivalent B\"uchi automata plays an
important role in many algorithms for LTL model checking, LTL satisfiability
checking etc. For a long time, researchers aimed to find fast translations
producing B\"uchi automata with a small number of states. This goal has led to
the developments of several translation algorithms and many heuristics and
optimizations including input formula reductions and optimizations of
produced B\"uchi automata, see
e.g.~\cite{Cou99,DGV99,EH00,SB00,GO01,GL02,Fri03,spot04}.

As the time goes, the translation objectives and their importance are
changing. In particular,~\cite{ST03} demonstrates that for higher
performance of the subsequent steps of the model checking process, it is
more important to minimize the number of states with nondeterministic choice
than the number of all states in resulting automata. Note that there are
LTL formulae, e.g.~$\F\G a$, for which no equivalent deterministic B\"uchi
automaton exists. Further, model checking practice shows that one LTL
formula is usually used in many different model checking tasks. Hence, it
pays to invest enough computation time to get high quality (more
deterministic and/or minimal) automata as it may reduce computation time of
many model checking tasks.

The new objectives lead to the developments of algorithms focusing
on quality of produced automata. For example,~\cite{DEK07} presents an
effective algorithm translating LTL formulae of the fragment called
\emph{obligation} (see~\cite{MP90}) into \emph{weak deterministic
  B\"uchi automata (WDBA)}. Moreover, WDBA can be minimized by the
algorithm of~\cite{Lod01}. There is also a SAT-based algorithm searching
for minimal (nondeterministic) B\"uchi automata~\cite{EF10}.
The main disadvantage of all the mentioned determinization and minimization
algorithms is their long running time which limits their use. 

\medskip Our research returns to the roots: we focus on a fast translation
producing a~relatively good output. This approach is justified by the
following facts:
\begin{itemize}
\item The mentioned algorithms producing high quality automata often need,
  for a given LTL formula, some equivalent automaton as an input.
\item The mentioned algorithms are usually feasible for short
  formulae only or for formulae with a simple structure.
\item Given a fresh LTL formula, it can be useful to run vacuity checks,
  namely satisfiability of the formula and its negation, to detect bugs in
  the formula. In these checks, time of the LTL to automata translation can
  be more significant than time needed for subsequent computations
  (see~\cite{RV07}). Hence, we need a fast translator to support an early
  detection of bugs in formulae.
\end{itemize}

Considering the speed of an LTL to B\"uchi automata translation,
LTL2BA~\cite{GO01} and SPOT~\cite{spot04} are two leading
tools. Based on extensive experiments on LTL satisfiability checking,
\cite{RV07} 
even states:
\begin{quotation}
  \emph{The difference in performance between SPOT and LTL2BA, on
    one hand, and the rest of explicit tools is quite dramatic.}
\end{quotation}
Each of the two tools is based on different algorithms. 

In LTL2BA, the translation proceeds in three basic steps:
\begin{enumerate}
\item A given LTL formula is translated into a \emph{very weak alternating
    automaton (VWAA)} with a co-B\"uchi accepting condition.
\item The alternating automaton is then translated into a
  \emph{transition-based generalized B\"uchi automaton (TGBA)}, i.e.~a
  generalized B\"uchi automaton with sets of accepting transitions instead
  of accepting states.
\item The generalized automaton is transformed (degeneralized) into a
  B\"uchi automaton (BA).
\end{enumerate}
Each of the three automata is simplified during the translation. 

SPOT translates a given LTL formula to a TGBA using a tableau method
presented in~\cite{Cou99}. The TGBA is then translated to a BA. Note that
the model checking algorithm natively implemented in SPOT works directly
with TGBAs.  Prior to a translation, both LTL2BA and SPOT try to
decrease the number of temporal operators in a given input formula by
applications of reduction rules.

While the LTL to automata translation in SPOT is under the gradual
development following the current trends (see~\cite{Lut11} for improvements
made in the last four years), LTL2BA underwent only one minor update in 2007
since its creation in 2001. In particular, SPOT reflects the changes in
objectives. Therefore, SPOT usually produces more deterministic and smaller
automata than LTL2BA, while LTL2BA is often a bit faster.

\paragraph{\textbf{Our contribution.}}
We introduce several modifications of~LTL2BA on both algorithmic and
implementation levels. We suggest changes in all the steps of the
translation algorithm. Our experimental results indicate that each modified
step has a mostly positive effect on the translation. The new translator,
called LTL3BA, is usually faster than the original LTL2BA and it produces
smaller and more deterministic automata. Moreover, comparison of LTL3BA and
the current version of SPOT (run without WDBA minimization that is very
slow) shows that the produced automata are of similar quality and LTL3BA is
usually faster.

Some modifications employ an observation that each LTL formula containing at
least one \emph{always} operator and at least one \emph{eventually} operator
on each branch of its syntax tree (with possible exceptions of branches
going to the left subformula of any \emph{until} or \emph{release} operator)
is prefix invariant. We call them \emph{alternating} formulae. Indeed,
validity of each alternating formula on a given word $u$ depends purely on a
suffix of $u$. In other words, it is not affected by any finite prefix of
$u$. We apply this observation to construct new rules for formula
reductions. Further, the observation justifies some changes in constructions
of VWAA and TGBA. Intuitively, a state of a VWAA corresponds to a subformula
that has to be satisfied by the rest of an accepted word. If the
corresponding subformula is an alternating formula, then the state can be
temporarily suspended for finitely many steps of the automaton.

Other changes in a VWAA construction are designed to lower
nondeterminism. This is also a motivation for new simplification rules
applied on intermediate automata. These rules remove some transitions of the
automaton and hence reduce the number of nondeterministic choices in
produced automata. The original simplification rules can be seen as special
cases of the new rules. An effective implementation of this simplification
required to change representation of transitions. Further, we add one ad-hoc
modification speeding up the translation of selected (sub)formulae. Finally,
we modify a simplification rule merging some states of resulting BA.

\medskip The rest of the paper is organized as follows. The next section
recalls the definitions of LTL, VWAA, and TGBA. Section~\ref{sec:alt}
focuses on alternating formulae and its
properties. Sections~\ref{sec:reduction},~\ref{sec:ltl2vwaa},
~\ref{sec:vwaa2tgba}, and~\ref{sec:tgba2ba} present new rules for formula
reductions, modified translation of LTL to VWAA (including generalized
simplification of VWAA), modified translation of VWAA to TGBA, and modified
rule for simplification of BA, respectively. Finally,
Section~\ref{sec:experiments} is devoted to experimental results. The last
section summarizes the achieved improvements.


\section{Preliminaries}\label{sec:prelim}

In this section, we recall the definition of LTL and definitions of VWAA and
TGBA as presented in~\cite{GO01}.

\subsubsection*{Linear Temporal Logic (LTL)}

The syntax of LTL~\cite{Pnu77} is defined as
follows
\[
\varphi~::=~\true~\mid~a~\mid~\neg\varphi~\mid~\varphi\vee\varphi~\mid~
\varphi\wedge\varphi~\mid~\X\varphi~\mid~ \varphi\U\varphi\textrm{,}
\]
where $\true$ stands for \emph{true}, $a$ ranges over a countable set $\AP$
of \emph{atomic propositions}, $\X$ and $\uU$ are temporal operators called
\emph{next} and \emph{until}, respectively. The logic is interpreted over
infinite words over the alphabet $\Sigma=2^{\AP'}$, where $\AP'\subseteq\AP$
is a finite subset. Given a word $u=u(0)u(1)u(2)\ldots\in\Sigma^\omega$,
by $u_i$ we denote the $i^{th}$ suffix of $u$, i.e.~$u_i=u(i)u(i+1)\ldots$.

The semantics of LTL formulae is defined inductively as follows:
\begin{center}
\begin{tabbing}
  \hspace*{1em} \= $u\models\true$\\
  \> $u\models a$ \hspace*{2.7em} \= iff~ \= $a\in u(0)$\\
  \> $u\models\neg\varphi$ \> iff \> $u\not\models\varphi$\\
  \> $u\models\varphi_1\vee\varphi_2$ \> iff \>
  $u\models\varphi_1$ or $u\models\varphi_2$\\
  \> $u\models\varphi_1\wedge\varphi_2$ \> iff \>
  $u\models\varphi_1$ and $u\models\varphi_2$\\
  \> $u\models\X\varphi$ \> iff \> $u_1\models\varphi$\\
  \> $u\models\varphi_1\U\varphi_2$ \> iff \>
     $\exists i\ge 0\,.\,(\,u_i\models\varphi_2$ and
     $\forall\, 0\leq j<i\,.~u_j\models\varphi_1\,)$
\end{tabbing}
\end{center}

We say that a word $u$ \emph{satisfies} $\varphi$ whenever
$u\models\varphi$. Two formulae $\varphi,\psi$ are \emph{equivalent},
written $\varphi\equiv\psi$, if for each alphabet $\Sigma$ and each
$u\in\Sigma^\omega$ it holds $u\models\varphi\iff u\models\psi$.
Given an alphabet $\Sigma$, a formula $\varphi$ defines
the language $L^\Sigma(\varphi)=\{u\in\Sigma^\omega\mid
u\models\varphi\}$. We often write $L(\varphi)$ instead of
$L^{2^{\AP(\varphi)}}(\varphi)$, where $\AP(\varphi)$ denotes the set of
atomic propositions occurring in the formula $\varphi$.

We extend the LTL with derived temporal operators:
\begin{itemize}
\item $\F\varphi$ called \emph{eventually} and equivalent to $\true\U\varphi$,
\item $\G\varphi$ called \emph{always} and equivalent to $\neg\F\neg\varphi$, and
\item $\varphi\R\psi$ called \emph{release} and equivalent to
  $\neg(\neg\varphi\,\U\neg\psi)$.
\end{itemize}

In the following, \emph{temporal formula} is a formula where the topmost
operator is neither conjunction, nor disjunction. A formula without any
temporal operator is called \emph{state formula}.  
Note that $a$ and $\true$ are both temporal and state formulae.  
An LTL formula is in \emph{positive normal form}
if no operator occurs in the scope of any negation. Each LTL formula can be
easily transformed to positive normal form using De Morgan's laws for
operators $\vee$ and $\wedge$, equivalences for derived operators, and the
following equivalences:
\[
\neg(\varphi_1\U\varphi_2) \equiv \neg\varphi_1\R\neg\varphi_2~~~~~~~~
\neg(\varphi_1\R\varphi_2) \equiv \neg\varphi_1\U\neg\varphi_2~~~~~~~~
\neg\X\varphi \equiv \X\neg\varphi 
\]


\subsubsection*{B\"uchi Automata (BA)}
A BA is a tuple $\mB=(Q,\Sigma,\delta,I,F)$, where
  \begin{itemize}
  \item $Q$ is a finite set of \emph{states},
  \item $\Sigma$ is a finite \emph{alphabet},
  \item $\delta: Q\rightarrow 2^{\Sigma\times Q}$ is a total
    \emph{transition function},
  \item $I\subseteq Q$ is a set of \emph{initial states}, and
  \item $F\subseteq Q$ is a set of \emph{accepting states}.
  \end{itemize}
%
Automaton $\mB$~is deterministic if and only if $|I| = 1$ and
$|\delta(q,a)|\leq 1$ for all $q\in Q$ and $a\in \Sigma$.

A \emph{run} $\rho$ of $\mB$ over an infinite word $w = w(0)w(1)w(2)\ldots
\in \Sigma^{\omega}$ is a sequence $\rho=q_0q_1q_2 \ldots$, where $q_0 \in
I$ is an initial state and $q_{i+1} \in \delta(q_{i}, w(i))$ for all $i \geq
0$. The run $\rho$ is \emph{accepting} if some accepting state occurs
infinitely often in the sequence $q_0q_1q_2 \ldots$. An infinite word $w$ is
\emph{accepted} by an automaton $\mB$ if some run of $\mB$ over $w$ is
accepting.

We denote by $L(\mB)$ the language accepted by $\mB$, i.e. the set of all
words over~$\Sigma$ accepted by an automaton $\mB$.


\subsubsection*{Very Weak Alternating co-B\"{u}chi Automata (VWAA)}
\label{def-VWAA}


A VWAA is a tuple $\mA = (Q, \Sigma, \delta, I, F)$, where
  \begin{itemize}
  \item $Q$ is a finite set of \emph{states}, and we let 
    $Q'=2^Q$,
  \item $\Sigma$ is a finite \emph{alphabet}, and we let $\Sigma'=2^\Sigma$,
  \item $\delta: Q \rightarrow 2^{\Sigma'\times Q'}$ is a \emph{transition
      function},
  \item $I \subseteq Q'$ is a set of \emph{initial states}, 
  \item $F \subseteq Q$ is a set of \emph{accepting states}, and
  \item there exists a partial order on $Q$ such that, for each state $q\in
    Q$, all the states occurring in $\delta(q)$ are lower or equal to $q$.
  \end{itemize}
  Note that the transition function $\delta$ uses $\Sigma'$ instead of
  $\Sigma$. This enables to merge transitions that differ only by action
  labels. We sometimes use a propositional formula $\alpha$ over $\AP$ to
  describe the element $\{a\in\Sigma\mid a\textrm{~satisfies~}\alpha\}$ of
  $\Sigma'$.

A \emph{run} $\sigma$ of VWAA $\mA$ over a word $w = w(0)w(1)w(2)\ldots \in
\Sigma^\omega$ is a labelled directed acyclic graph $(V,E,\lambda)$ such
that:
\begin{itemize}
\item $V$ is partitioned into $\bigcup^{\infty}_{i=0}\limits V_i$ with
  $E\subseteq \bigcup^{\infty}_{i=0} \limits V_i\times V_{i+1}$,
\item $\lambda: V \rightarrow Q$ is a labelling function,
\item $\{\lambda(x)\mid x \in V_0\}\in I$, and
%
\item for each $x \in V_i$, there exist $\alpha\in\Sigma'$, $q \in Q$ and $O
  \in Q'$ such that $w(i)\in\alpha$, $q =\lambda(x)$, $O = \{\lambda(y) \mid
  (x,y)\in E\}$, and $(\alpha, O) \in \delta(q)$.

\end{itemize}
A run $\sigma$ is \emph{accepting} if each branch in $\sigma$ contains only
finitely many nodes labelled by accepting states (co-B\"uchi acceptance
condition). A word $w$ is \emph{accepted} if there is an accepting run over
$w$.

We denote by $L(\mA)$ the language accepted by $\mA$, i.e. the set of all
words over~$\Sigma$ accepted by an automaton $\mA$. 

\subsubsection*{Transition Based Generalized B\"{u}chi Automata (TGBA)}

  A TGBA is a
  tuple $\mG=(Q,\Sigma,\delta, I, \mF)$, where
  \begin{itemize}
  \item $Q$ is a finite set of \emph{states},
  \item $\Sigma$ is a finite \emph{alphabet}, and we let $\Sigma'=2^\Sigma$
  \item $\delta: Q \rightarrow 2^{\Sigma' \times Q}$ is a total
    \emph{transition function}, 
  \item $I \subseteq Q$ is a set of \emph{initial states}, and
  \item $\mT = \{T_1, T_2, \ldots, T_m\}$ where $T_j \subseteq Q \times
    \Sigma' \times Q$ are sets of \emph{accepting transitions}.
  \end{itemize}

A run $\rho$ of TGBA $\mG$ over a word $w = w(0)w(1)w(2)\ldots \in
\Sigma^{\omega}$ is a sequence of states $\rho=q_0q_1q_2 \ldots, $ where
$q_0 \in I$ is an initial state and, for each $i\geq 0$, there exists
$\alpha\in\Sigma'$ such that $w(i) \in \alpha$ and
$(\alpha,q_{i+1})\in\delta(q_{i})$. A run $\rho$ is \emph{accepting} if for
each $1 \leq j \leq m$ it uses infinitely many transitions from $T_j$.
A word $w$ is \emph{accepted} if there is an accepting run over
$w$.

We denote by $L(\mG)$ the language accepted by $\mG$, i.e. the set of all
words over~$\Sigma$ accepted by an automaton $\mG$.


\section{Alternating Formulae}\label{sec:alt}


We define the class of \emph{alternating formulae} together with the classes
of \emph{pure eventuality} and \emph{pure universality} formulae introduced
in~\cite{EH00}. Let $\varphi$ ranges over general LTL formulae. The
following abstract syntax equations define the classes \emph{pure
  eventuality} formulae $\mu$, \emph{pure universality} formulae $\nu$, and
\emph{alternating} formulae $\xi$:
\[
\begin{array}{rcl}
  \mu&::=&\F\varphi~\mid~\mu\vee\mu~\mid~\mu\wedge\mu
  ~\mid~\X\mu~\mid~\varphi\U\mu
  ~\mid~\mu\R\mu~\mid~\G\mu\\[.7ex]
  \nu&::=&\G\varphi~\mid~\nu\vee\nu~\mid~\nu\wedge\nu
  ~\mid~\X\nu~\mid~\nu\U\nu~\mid~\varphi\R\nu~\mid~\F\nu\\[.7ex]
  \xi&::=&\G\mu~\mid~\F\nu~\mid~\xi\vee\xi~\mid~
  \xi\wedge\xi~\mid~\X\xi~\mid~\varphi\U\xi~\mid~
  \varphi\R\xi~\mid~\F\xi~\mid~\G\xi
\end{array}
\]
Note that there are alternating formulae, e.g.~$\big(a\U(\G\F
b)\big)\wedge\big(c\R(\G\F d)\big)$, that are neither pure eventuality
formulae, nor pure universality formulae. Properties of the respective
classes of formulae are summarized in the following lemmata.
\begin{lemma}{\cite{EH00}}
  \label{lem:pure}
  Every pure eventuality formula $\mu$ satisfies the following:
  \[
    \forall w\in\Sigma^{\omega}, u\in\Sigma^*: 
    w\models\mu\implies uw\models\mu
  \]
  Further, every pure universality formula $\nu$ satisfies the following:
  \[
    \forall w\in\Sigma^{\omega}, u\in\Sigma^*: 
    uw\models\nu\implies w\models\nu
  \]
\end{lemma}
In other words, pure eventuality formulae define \emph{left-append closed}
languages while pure universality formulae define \emph{suffix closed}
languages.
\begin{lemma}\label{lem:alt}
  Every alternating formula $\xi$ satisfies the following:
  \[
    \forall w\in\Sigma^{\omega}, u\in\Sigma^*: 
    uw\models\xi\iff w\models\xi
  \]
\end{lemma}
In other words, each alternating formula defines a \emph{prefix-in\-variant}
language. 
\begin{proof}
  The proof proceeds by induction on the structure of $\xi$. We assume that
  $w\in\Sigma^{\omega}$ is an arbitrary infinite word and $u\in\Sigma^*$ is
  an arbitrary finite word.
  \begin{description}
  \item[$\xi=\G\mu$] --~ The semantics of $\G$ operator directly provides one
    implication, namely $uw\models\G\mu \implies w\models\G\mu$. As $\mu$ is
    a pure eventuality formula, Lemma~\ref{lem:pure} gives us $\forall
    u'\in\Sigma^*:w\models\mu \implies
    u'w\models\mu$. This implies $w\models\G\mu \implies
    uw\models\G\mu$. In total, we get $w\models\G\mu \iff uw\models\G\mu$.
  \item[$\xi=\F\nu$] --~ The semantics of $\F$ operator directly provides one
    implication, namely $w\models\F\nu \implies uw\models\F\nu$. As $\nu$ is
    a pure universality formula, Lemma~\ref{lem:pure} gives us $\forall
    u'\in\Sigma^*:u'w\models\nu \implies w\models\nu$. This implies
    $uw\models\F\nu \implies w\models\F\nu$. In total, we get $w\models\F\nu
    \iff uw\models\F\nu$.
  \item[$\xi=\varphi\U\xi_1$] --~ From the induction hypothesis, it follows
    that $\F\xi_1\implies\xi_1$. Hence, $\varphi\U\xi_1 \equiv \xi_1$
    holds. Thus, the statement coincides with the induction hypothesis.
  \item[$\xi=\varphi\R\xi_1$] --~ From the induction hypothesis, it follows
    that $\xi_1\implies\G\xi_1$ and thus also
    $\xi_1\implies\varphi\R\xi_1$. As $\varphi\R\xi_1\implies\xi_1$, we get
    $\varphi\R\xi_1\equiv\xi_1$. Hence, the statement coincides with the
    induction hypothesis.
  \item[$\xi=\xi_1\vee\xi_2$ or $\xi=\xi_1\wedge\xi_2$ or $\xi=\X\xi_1$ or
    $\xi=\F\xi_1$ or $\xi=\G\xi_1$] --~ In all these cases, the statement
    easily follows from the induction hypothesis. \qed
 \end{description}
\end{proof}

\begin{corollary}
  Every alternating formula $\xi$ satisfies $\xi \equiv \X\xi$.
\end{corollary}
Hence, in order to check whether $w$ satisfies $\xi$ it is possible to skip
an arbitrary long finite prefix of the word $w$.

We use this property in new rule for formula reduction. Further, it has
brought us to the notion of alternating formulae \emph{suspension} during
the translation of LTL to B\"{u}chi automata. We employ suspension on two
different levels of the translation: the construction of a VWAA from an input
LTL formula and the transformation of a VWAA into a TGBA.


\section{Improvements in Reduction of LTL Formulae}
\label{sec:reduction}

Many rules reducing the number of temporal operators in an LTL
formula have been presented in~\cite{SB00} and~\cite{EH00}. In this section
we present some new reduction rules. For the rest of this section,
$\varphi,\psi$ range over LTL formulae and $\gamma$ ranges over
alternating ones.
\[
\begin{array}{rclp{4ex}rclp{4ex}rclp{4ex}rcl}
  \X \varphi \R \X \psi &\equiv& \X (\varphi \R \psi) && 
  \varphi \U \gamma &\equiv& \gamma &&
  \F \gamma &\equiv& \gamma &&
  \X \gamma &\equiv& \gamma \\
  \X \varphi \vee \X \psi &\equiv& \X (\varphi \vee \psi) &&
  \varphi \R \gamma &\equiv& \gamma &&
  \G \gamma &\equiv& \gamma &&   & & 
\end{array}
\]
The following equivalences are valid only on assumption that 
$\varphi$ implies $\psi$.
\[
\begin{array}{rclp{4ex}rcl}
  \psi \U (\varphi \U \gamma) &\equiv& \psi \U \gamma &&
  \varphi \wedge (\psi \wedge \gamma) &\equiv& (\varphi \wedge \gamma)\\
  (\psi \R \gamma) \R \varphi &\equiv& \gamma \R \varphi &&
  \psi \vee (\varphi \vee \gamma) &\equiv& (\psi \vee \gamma)\\
  \varphi \U (\gamma \R (\psi \U \rho)) &\equiv& \gamma \R (\psi \U \rho)
\end{array}
\]
Further, we have extended the set of rules deriving implications of the form
$\varphi \Rightarrow \psi$. The upper formula is a precondition, the lower
one is a conclusion.
\[
\begin{array}{cp{4ex}cp{4ex}c}
  \dfrac{\G \varphi \Rightarrow \psi}{\G \varphi \Rightarrow \X \psi} && 
  \dfrac{\varphi \Rightarrow \F \psi}{\X \varphi \Rightarrow \F \psi} && 
  \dfrac{\varphi \Rightarrow \psi}{\X \varphi \Rightarrow \X \psi}
\end{array}
\]



\section{Improvements in LTL to VWAA Translation}
\label{sec:ltl2vwaa}

First, we recall the original translation of LTL to VWAA according
to~\cite{GO01}.
%
%
%
The translation utilizes two auxiliary operators:
\begin{itemize}
\item Let $\Sigma' = 2^{\Sigma}$, and let $Q' = 2^Q$.
  Given $J_1,J_2\in 2^{\Sigma' \times Q'}$, we
  define $$J_1 \otimes J_2=\{(\alpha_1 \cap \alpha_2, O_1 \cup O_2) \mid
  (\alpha_1, O_1) \in J_1 \mbox{ and } (\alpha_2, O_2) \in J_2\}.$$
\item Let $\psi$ be an LTL formula in positive normal form. We define
  $\overline{\psi}$ by:
  \begin{itemize}
  \item$\overline{\psi}=\{\{\psi\}\}$ if $\psi$ is a temporal formula,
  \item $\overline{\psi_1 \wedge \psi_2} = \{O_1 \cup O_2 \mid O_1 \in
    \overline{\psi_1}\ \mbox{and}\ O_2 \in \overline{\psi_2}\}$,
  \item $\overline{\psi_1 \vee \psi_2} = \overline{\psi_1} \cup
    \overline{\psi_2}$.
  \end{itemize}
\end{itemize}


Let $\varphi$ be an LTL formula in positive normal form. An equivalent VWAA
with a co-B\"{u}chi acceptance condition is constructed as $\mA_\varphi=(Q,
\Sigma, \delta, I, F)$, where $Q$ is the set of temporal subformulae of
$\varphi$, $\Sigma=2^{\AP(\varphi)}$, $I=\overline{\varphi}$, $F$ is the set
of all $\uU$-subformulae of $\varphi$, i.e formulae of the type
$\psi_1\U\psi_2$, and $\delta$ is defined as follows:
\[
\begin{array}{rcl}
  \delta(\true) & = & \{(\Sigma,\emptyset)\} \\
  \delta(p) & = & \{(\Sigma_p, \emptyset)\}\ 
  \textrm{where}\ \Sigma_p = \{a\in\Sigma\mid p \in a\} \\
  \delta(\neg p) & = & \{(\Sigma_{\neg p}, \emptyset)\}\ 
  \textrm{where}\ \Sigma_{\neg p} = \Sigma\smallsetminus\Sigma_p \\
  \delta(\X\psi) & = & \{(\Sigma,O)\mid O\in\overline{\psi}\} \\
  \delta(\psi_1\U\psi_2) & = & \Delta(\psi_2)\cup
  \big(\Delta(\psi_1)\otimes\{(\Sigma,\{\psi_1\U\psi_2\})\}\big) \\
  \delta(\psi_1\R\psi_2) & = & \Delta(\psi_2)\otimes
  \big(\Delta(\psi_1)\cup\{(\Sigma,\{\psi_1\R\psi_2\})\}\big) \\
  \\
  \Delta(\psi) & = & 
  \delta(\psi)\ \textrm{if}\ \psi\ \textrm{is a temporal formula} \\
  \Delta(\psi_1\vee\psi_2) & = & \Delta(\psi_1)\cup\Delta(\psi_2) \\
  \Delta(\psi_1\wedge\psi_2) & = & \Delta(\psi_1)\otimes\Delta(\psi_2) \\
\end{array}
\]
Using the partial order ``is a subformula of'' on states of $\mA_\varphi$,
one can easily prove that $\mA_\varphi$ is very weak.

\subsubsection*{Improved Translation}

In order to implement the suspension of alternating formulae, we modify the
way the transition function $\delta$ handles the binary operators $\uU$,
$\rR$, $\vee$, and $\wedge$. The original transition function $\delta$
reflects the following identities:
\[
\begin{array}{rcl}
  \varphi_1 \U \varphi_2 & \equiv &
  \varphi_2 \vee (\varphi_1 \wedge \X (\varphi_1 \U \varphi_2)) \\
  \varphi_1 \R \varphi_2 & \equiv &
  \varphi_2 \wedge (\varphi_1 \vee \X (\varphi_1 \R \varphi_2))
\end{array}
\]
However, if $\varphi_1$ is an alternating formula we apply the relation
$\varphi_1 \equiv \X \varphi_1$ to obtain the following identities:
\[
\begin{array}{rcl}
  \varphi_1 \U \varphi_2 & \equiv &
  \varphi_2 \vee (\X \varphi_1 \wedge \X (\varphi_1 \U \varphi_2)) \\
  \varphi_1 \R \varphi_2 & \equiv &
  \varphi_2 \wedge (\X \varphi_1 \vee \X (\varphi_1 \R \varphi_2))
\end{array}
\]
Using these identities, the formula $\varphi_1$ is effectively suspended and
checked one step later. Similarly, in the case of disjunction or
conjunction, each disjunct or conjunct corresponding to an alternating
formula is suspended for one step as well. Correctness of these changes
clearly follows from properties of alternating formulae.
Note that $\delta$ is defined over formulae in positive normal form
only. The translation treats each formula $\F\psi$ as $\true\U\psi$ and each
formula $\G\psi$ as $(\neg \true) \R \psi$.

We introduce further changes to the transition function $\delta$ in order to
generate automata which exhibits more determinism. In particular, we build a
VWAA with only one initial state. Similarly, each state corresponding to a
formula of a type $\X\varphi$ generates only one successor corresponding to
$\varphi$. These changes can add an extra initial state and an extra state
for each $\X$-subformula comparing to the original construction. However,
this drawback is often suppressed due to the consecutive optimizations
during the construction of a TGBA.

Now we present a modified construction of VWAA. Given an input LTL formula
$\varphi$ in positive normal form, an equivalent VWAA with a co-B\"uchi
acceptance condition is constructed as $\mA_\varphi=(Q,\Sigma,\delta,I,F)$,
where $Q$ is the set of all subformulae of $\varphi$, $\Sigma$ and $F$
are defined as in the original construction, $I = \{\varphi\}$, and $\delta$
is defined as follows:
%
\[
\begin{array}{rcl}
  \delta(\true) & = & \{(\Sigma,\emptyset)\} \\
  \delta(p) & = & \{(\Sigma_p, \emptyset)\}\ 
  \textrm{where}\ \Sigma_p = \{a\in\Sigma\mid p \in a\} \\
  \delta(\neg p) & = & \{(\Sigma_{\neg p}, \emptyset)\}\ 
  \textrm{where}\ \Sigma_{\neg p} = \Sigma\backslash\Sigma_p \\
  \delta(\X\psi) & = & \{(\Sigma,\{\psi\})\} \\
  \delta(\psi_1\vee\psi_2) & = & \Delta(\psi_1)\cup\Delta(\psi_2) \\
  \delta(\psi_1\wedge\psi_2) & = & \Delta(\psi_1)\otimes\Delta(\psi_2) \\
  \delta(\psi_1\U\psi_2) & = & \left\lbrace
    \begin{array}{ll}
      \Delta(\psi_2)\cup
      (\{(\Sigma,\{\psi_1\})\}\otimes\{(\Sigma,\{\psi_1 \U \psi_2\})\})
      & \textrm{if } \psi_1 \textrm{ is alternating,} \\
      \Delta(\psi_2) \cup
      (\Delta(\psi_1)\otimes\{(\Sigma,\{\psi_1\U\psi_2\})\})
      & \textrm{otherwise.} \\
    \end{array}
  \right. \vspace{0.2em} \\
  \delta(\psi_1\R\psi_2) & = & \left\lbrace
    \begin{array}{ll}
      \Delta(\psi_2)\otimes
      (\{(\Sigma, \{\psi_1\}), (\Sigma, \{\psi_1 \R \psi_2\})\})
      & \textrm{if } \psi_1 \textrm{ is alternating,} \\
      \Delta(\psi_2)\otimes
      (\Delta(\psi_1)\cup\{(\Sigma,\{\psi_1\R\psi_2\})\})
      & \textrm{otherwise.} \\		
    \end{array}
  \right. \\
  \\
  \Delta(\psi) & = &  \left\lbrace
    \begin{array}{ll}
      \{(\Sigma,\{\psi\})\}~ & 
      \textrm{if}\ \psi\ \textrm{is a temporal alternating formula,} \\[1ex]
      \delta(\psi) & \textrm{if}\ \psi\ \textrm{is a temporal
        formula that is not alternating.} 
    \end{array}
  \right. \\
  \Delta(\psi_1\vee\psi_2) & = & \Delta(\psi_1)\cup\Delta(\psi_2) \\
  \Delta(\psi_1\wedge\psi_2) & = & \Delta(\psi_1)\otimes\Delta(\psi_2) \\
	\end{array}
	\]


\begin{figure}[tb]
  \[
  \begin{array}[b]{rcrcrc}
    (a) &
    \xymatrix@C=1.5ex@R=2.9ex{
      & \ar[d] \\
      & *+[F=:<5pt>]{1: (\G\F a) \U b} 
      \ar@/_1.7pc/@<-2.2ex>[dddd]
      \ar@(dl,d)@<-2.3ex>[] \ar@/_1.2 pc/@<-1.75ex>[dd]_(.4){\true} 
      \ar@(dr,d)@<+2.3ex>[] \ar@/^1.2pc/@<+1.75ex>[dd]^(.4){a} 
      \ar[rr]^(.7){b}
      && \\ \\
      & *+[F-:<5pt>]{2: \G\F a} 
      \ar@(dl,d)_(.25){a\!}
      \ar@/^1.2pc/@<+1.55ex>[dd]^(.4){\true} \ar@(dr,d)@<+2.1ex>[] 
      \\ \\
      & *+[F=:<5pt>]{3: \F a} \ar@(d,dl)^(.25){\true} \ar[r]_(0.6){a} & \\
    } 
    & ~~~~~(b) &
    \xymatrix@C=1.5ex@R=2.9ex{
      & \ar[d] \\
      & *+[F=:<5pt>]{1: (\G\F a) \U b} 
      \ar@(dl,d)@<-2.3ex>[] \ar@/_1.2 pc/@<-1.75ex>[dd]_(.4){\true} 
      \ar[rr]^(.7){b}
      && \\ \\
      & *+[F-:<5pt>]{2: \G\F a} 
      \ar@(dl,d)_(.25){a}
      \ar@/^1.2pc/@<+1.55ex>[dd]^(.4){\true} \ar@(dr,d)@<+2.1ex>[] 
      \\ \\
      & *+[F=:<5pt>]{3: \F a} \ar@(d,dl)^(.25){\true} \ar[r]_(0.6){a} & \\
    } 
    & ~~~~~(c) &
    \xymatrix@C=1.5ex@R=2.9ex{
      & \ar[d] \\
      & *+[F=:<5pt>]{1: (\G\F a) \U b} 
      \ar@(dl,d)@<-2.3ex>[] \ar@/_1.2 pc/@<-1.75ex>[dd]_(.4){\neg b} 
      \ar[rr]^(.7){b}
      && \\ \\
      & *+[F-:<5pt>]{2: \G\F a} 
      \ar@(dl,d)_(.25){a}
      \ar@/^1.2pc/@<+1.55ex>[dd]^(.4){\neg a} \ar@(dr,d)@<+2.1ex>[] 
      \\ \\
      & *+[F=:<5pt>]{3: \F a} \ar@(d,dl)^(.25){\neg a} \ar[r]_(0.6){a} & \\
    } 
  \end{array}
  \]
  \caption{VWAA for $(\G\F a) \U b$ generated by (a) the translation
    of~\cite{GO01}, (b) our translation with suspension, and (c) our
    translation with suspension and further determinization.}
  \label{fig:alt}
\end{figure}

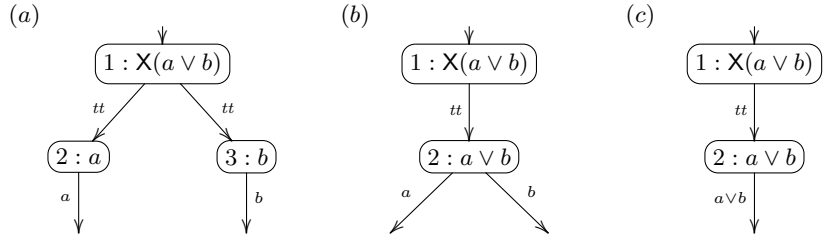
\begin{figure}[tb]
  \[
  \begin{array}[b]{rcrcrc}
    (a) &
    \xymatrix@!C=2ex@R=2ex{
      & \ar[d] \\
      & *+[F-:<5pt>]{1: \X(a\vee b)} 
      \ar[ddl]_(.6)\true \ar[ddr]^(.6)\true \\ \\
      *+[F-:<5pt>]{2: a} \ar[dd]_a 
      && *+[F-:<5pt>]{3: b} \ar[dd]^b 
      \\ \\
      &&
    } 
    & ~~~~~~~(b) &
    \xymatrix@!C=2ex@R=2ex{
      & \ar[d] \\
      & *+[F-:<5pt>]{1: \X(a\vee b)} 
      \ar[dd]_\true \\ \\
      & *+[F-:<5pt>]{2: a\vee b} \ar[ddl]_(.6)a \ar[ddr]^(.6)b 
      \\ \\
      &&
    } 
    & ~~~~~~~(c) &
    \xymatrix@!C=2ex@R=2ex{
      & \ar[d] \\
      & *+[F-:<5pt>]{1: \X(a\vee b)} 
      \ar[dd]_\true \\ \\
      & *+[F-:<5pt>]{2: a\vee b} \ar[dd]_{a\vee b}
      \\ \\
      &&
    } 
  \end{array}
  \]
  \caption{VWAA for $\X(a\vee b)$ generated by (a) the translation
    of~\cite{GO01}, (b) our translation with suspension, and (c) the
    translation with suspension and further determinization.}
  \label{fig:X}
\end{figure}

Motivation for our changes in the translation can be found in
Figures~\ref{fig:alt} and~\ref{fig:X}. Each figure contains (a) the VWAA
constructed by the original translation and (b) the VWAA constructed by our
translation with suspension. Figure~\ref{fig:alt} shows the effect of
suspension of alternating subformula $\G\F a$ in computation of transitions
leading from the initial state. It can be easily proved that whenever one
start with a formula reduced according to Section~\ref{sec:reduction}, then
each suspension of an alternating temporal subformula leads just to
reduction of transitions in the resulting VWAA, i.e., no state is added. On
the other hand, if an alternating non-temporal subformula $\psi$ is
suspended or the new definition of $\delta(\X\psi)$ is used, then the resulting
VWAA can contain one more reachable state corresponding to the formula
$\psi$. However, other states may become unreachable and, in particular, the
automaton can also have more deterministic states as illustrated by
Figure~\ref{fig:X}.

\subsubsection*{Optimization of VWAA}

In the original algorithm, the VWAA is optimized before it is translated to
a TGBA. In particular, if there are two transitions $t_1=(q,\alpha_1,O_1)$
and $t_2=(q,\alpha_2,O_2)$ satisfying $\alpha_2\subseteq\alpha_1$ and
$O_1\subseteq O_2$, then $t_2$ is removed as it is implied by $t_1$.

We suggest a generalization of this principle: if $O_1\subsetneq O_2$ then
replace the label $\alpha_2$ in $t_2$ by $\alpha_2\wedge\neg\alpha_1$. If
$O_1=O_2$, replace both transitions by the transition $(q,
\alpha_1\vee\alpha_2, O_1)$.  Note that if $\alpha_2\Rightarrow\alpha_1$,
i.e.~$\alpha_2 \subseteq \alpha_1$, then $\alpha_2\wedge\neg\alpha_1 \equiv
\neg \true$ and transition $t_2$ can be removed as before. Our generalized
optimization rule increase determinism of the produced VWAA as illustrated
by automata (c) of Figures~\ref{fig:alt} and ~\ref{fig:X}.


\section{Improvements in VWAA to TGBA Translation}
\label{sec:vwaa2tgba}
 
First, we recall the translation of VWAA to TGBA introduced
in~\cite{GO01}. Let $\mA_\varphi = (Q, \Sigma, \delta, I, F)$ be a VWAA with
a co-B\"{u}chi acceptance condition.  We define $\mG_\mA = (Q', \Sigma,
\delta', I, \mathcal{T})$ to be a TGBA where:
\begin{itemize}
\item $Q' = 2^Q$, i.e.~a state is a set of states of $\mA_\varphi$ and
  represents their conjunction, 
\item $\delta'' (\{q_1, q_2, \ldots, q_n\}) = \bigotimes^n_{i=1}\limits
  \delta(q_i)$ is the non-optimized transition function,
\item $\delta'$ is the optimized transition function defined as the set of
  $\preccurlyeq$-minimal transitions of $\delta''$ where the relation
  $\preccurlyeq$ is defined by $t_1\preccurlyeq t_2$ iff $t_1 = (O,
  \alpha_1, O_1)$, $t_2 = (O, \alpha_2, O_2)$, $\alpha_2 \subseteq
  \alpha_1$, $O_1 \subseteq O_2$, and $\forall T_f \in \mathcal{T}$, $t_2
  \in T_f \Rightarrow t_1 \in T_f$, and
\item $\mathcal{T} = \{T_f \mid f \in F\}$ where \\
  $T_f = \{(O, \alpha, O') \mid f \not\in O'\ \textrm{or}\ \exists (\beta,
  O'') \in \delta(f), \alpha \subseteq \beta\ \textrm{and}\ f \not\in O''
  \subseteq O'\}$.
\end{itemize}

\subsubsection*{Improved Translation}

Our algorithm for a VWAA to TGBA translation differs from the original one
only in definition of $\delta$, where we also integrate the idea of
suspension of alternating formulae. Recall that each state $q_i$ of a VWAA
is a subformula of an input LTL formula and each state of a TGBA is
identified with a conjunction of states of a VWAA. Let $O = \{ q_1, 
\ldots, q_n\}$ be a state of a TGBA. Then transitions leading from $O$
in a TGBA correspond to combinations of transitions leading from
$q_1,\ldots,q_n$ in a VWAA. If $q_i$ is an alternating formula and thus it
satisfies $q_i\equiv\X q_i$, we can effectively decrease the number of
transition combinations that need to be considered during computation of
$\delta'(O)$ provided we suspend a full processing of $q_i$ to the succeeding
states of the TGBA. More precisely, for the purpose of computation of
$\delta'(O)$, we set $\delta(q_i)=\{(\Sigma,\{q_i\})\}$. To construct a TGBA
equivalent to the VWAA, we have to ensure that $q_i$ will not be suspended
forever during any accepting run of the TGBA. Hence, we enable suspension
only in the states that are not on any accepting cycle in a TGBA.

Let $M$ be the minimal set containing all VWAA states of the form
$\psi\R\rho$ and all subformulae of their right operands $\rho$. One can
observe
 each TGBA state lying on some accepting cycle is a subset of
$M$. The VWAA states outside $M$, called \emph{progress formulae}, push TGBA
computations towards accepting cycles. Suspension is enabled in a TGBA state
only if it contains a progress formula. However, if all progress formulae in
a TGBA state are alternating, their suspension is not allowed (as suspended
progress formulae would not enforce any progress).



Formally, for each TGBA state $O=\{q_1, q_2,\ldots, q_n\}$ we define
$\delta''(O)$ as follows:
$$\delta''(O) = 
\bigotimes^n_{i=1}\limits\delta_{O}(q_i) \textrm{, where}$$
\[ \delta_O(q_i) = \left\lbrace
  \begin{array}{ll}
    \{(\Sigma,\{q_i\})\}~~ 
    & \textrm{if $O$ contains a progress non-alternating formula}\\
    & \textrm{and $q_i$ is an alternating formula,}\\
    & \textrm{or $O$ contains a progress formula}\\
    & \textrm{and $q_i$ is an alternating non-progress formula,} \\[2ex]
    \delta(q_i) & \textrm{otherwise.}
  \end{array} \right.
\]
We have obtained better results when we restrict the definition of progress
formulae to temporal progress formulae.

Note that the original translation of VWAA to TGBA uses a correct but
non-intuitive definition of accepting sets $T_f$. In fact, our modification is
correct only if we change the definition of these sets to intuitive
one: for each accepting state $f$ of the VWAA with a co-B\"uchi
acceptance, we compute a set $T_f$ to contain all TGBA transitions that do
not contain any VWAA transition looping in $f$. Formally, $\mathcal{T} =
\{T_f \mid f \in F\}$ where
\[
  \setlength{\arraycolsep}{0pt}
  \begin{array}{rl}
    T_f = \{(O, \alpha, O') \mid f \not\in O'
    \textrm{~or~}&(\exists (\beta,O'')\in\delta(f), \exists 
    (\gamma,O''')\in 
  \bigotimes_{f'\in O\smallsetminus\{f\}} \delta(f') \\
  & \textrm{~such that~}
  f\not\in O''\textrm{, }
  \alpha=\beta\wedge\gamma\textrm{,~and~}O'=O''\cup O''')\}.
  \end{array}
\]
\begin{figure}[tb]
  \begin{minipage}[b]{0.45\linewidth}
    \[\xymatrix@C=2ex@R=2ex{
      &&&*+[F-:<5pt>]{1: \G\F a} \ar@(ul,ur)^{a} \ar@(dr,r)_{}
      \ar@(dr,dl)[rr]^(.5){\true} &&
      *+[F=:<5pt>]{3: \F a} \ar@(ul,ur)^{\true} \ar[dr]^(0.6){a}\\
      \ar@/_0.5pc/[rrru] \ar@/^0.5pc/[rrrd] &&&&&&&\\
      &&& *+[F=:<5pt>]{2: \F b} \ar@(dr,dl)^{\true} \ar[dr]^(0.6){b} \\
      &&&& }\]
    \caption{A VWAA $\mA_\psi$ corresponding to $\G\F a~\wedge~\F b$.}
    \label{fig:const_ex-alt}
  \end{minipage}
  \hfill
  \begin{minipage}[b]{0.45\linewidth}
    \[\xymatrix@C=4ex@R=2ex{
      \ar[r] & *+[F-:<5pt>]{\{1,2\}} \ar@(ul,u)^(.7){\true:\emptyset}
      \ar[rr]^{b:\{2,3\}} && *+[F-:<5pt>]{\{1\}} \ar@(ul,u)^(.7){a:\{2,3\}}
      \ar@(dr,d)^(.7){\true:\{2\}} }\]
    \caption{A TGBA $\mG_\psi$ corresponding to the VWAA of
      Figure~\ref{fig:const_ex-alt}.}
    \label{fig:const_ex-gen_full}
  \end{minipage}
\end{figure}

\begin{figure}[p]
  \begin{minipage}[t]{\linewidth}
    \[\xymatrix@C=4ex@R=3ex{
      &&*+[F-:<5pt>]{1: \G\F q} \ar@(ul,ur)^{q} \ar@(dr,r)_{}
      \ar@(dr,dl)[rr]^(.5){\true} &&
      *+[F=:<5pt>]{3: \F q} \ar@(ul,ur)^{\true} \ar[dr]^(0.6){q} \\
      \ar@/_0.5pc/[rru] \ar@/^0.5pc/[rrd] &&&&&&& \\
      && *+[F=:<5pt>]{2: \psi} \ar@(dr,r)_{} \ar@(dr,dl)[rr]^(.5){\true}
      \ar[dl]_(0.6){p_1} \ar@(d,dl)^{} \ar[dd]_(.6){\true} &&
      *+[F=:<5pt>]{4: \neg p_1\U p_3}  \ar@(ul,ur)^{\neg p_1} \ar[dr]^(0.7){p_3} \\
      &&&&&& \\
      &&*+[F-:<5pt>]{5: p_1\R p_2} \ar@(dr,dl)^{p_2} \ar[dr]^(0.7){p_1\wedge p_2} \\
      &&&&}\] 
    \caption{A VWAA $\mA_\varphi$ corresponding to formula $\varphi=\psi\wedge\G\F q$,
    where $\psi=(\X((p_1\R p_2)\vee(\neg p_1\U p_3)))\U p_1$.}
    \label{fig:fin_ex-alt}
  \end{minipage}
  \begin{minipage}[h]{\linewidth}
    \[\xymatrix@C=2.5ex@R=3ex{
    	\\ &&&&& \\
      & \ar[r] & *+[F-:<5pt>]{~\{1,2\}~} \ar[rrrr]^{\true:\{3\}}
      \ar[rrdd]^(.35){\true:\{3,4\}} \ar@/_/[dddd]^{p_1:\{2,3,4\}} && &&
      *+[F-:<5pt>]{\{1,2,4\}} \ar@<.5ex>@(u,ur)^(.7){\neg p_1:\{3\}} 
      \ar@[red]@/^/[lldd]^{\color{red}p_3:\{3,4\}} \ar@<.1ex>@(dr,d)^(.3){p_3:\{3,4\}}
      \ar `u[ul] `l[llllll]_{p_1\wedge p_3:\{2,3,4\}} `d[dddd] [lllldddd] \\ \\
      && && *+[F-:<5pt>]{\{1,2,5\}} \ar@(ul,u)^(.6){p_2:\{3,4\}}
      \ar[lldd]^(.45){p_1\wedge p_2:\{2,3,4\}} \ar@/_/[rrdd]_(.55){p_2:\{3\}}
      \ar@[red]@/^/[rruu]^(.6){\color{red}p_1\wedge p_2:\{2,3\}} \\ \\
      && *+[F-:<5pt>]{~\{1\}~} \ar@(u,ur)^(.5){\true:\{2,4\}}
      \ar@(dr,d)^(.7){q:\{2,3,4\}} && &&
      *+[F-:<5pt>]{\{1,2,4,5\}} \ar[llll]^{p_1\wedge p_2\wedge p_3:\{2,3,4\}}
      \ar@/_/[lluu]_(.4){p_2\wedge p_3:\{3,4\}} \ar@(dr,d)^(.7){\neg p_1\wedge p_2:\{3\}} }\]
    \caption{A TGBA $\mG_\varphi$ corresponding to the VWAA of
      Figure~\ref{fig:fin_ex-alt} constructed using original definition of accepting sets.}
    \label{fig:fin_ex-gen}
  \end{minipage}
  \begin{minipage}[b]{\linewidth}
  \[\xymatrix@C=2.5ex@R=3ex{
    	\\ &&&&&& \\
      && \ar[r] & *+[F-:<5pt>]{~\{1,2\}~} \ar[rrrr]^{\true:\{3\}}
      \ar[rrdd]^(.35){\true:\{3,4\}} \ar@/_/[dddd]^{p_1:\{2,3,4\}} && &&
      *+[F-:<5pt>]{\{1,2,4\}} \ar@<.5ex>@(u,ur)^(.7){\neg p_1:\{3\}} 
      \ar@/^/[lldd]^{p_3:\{3,4\}} \ar@<.1ex>@(dr,d)^(.3){p_3:\{3,4\}}
      \ar `u[ul] `l[llllll]_{p_1\wedge p_3:\{2,3,4\}} `d[dddd] [lllldddd] \\ \\
      &&& && *+[F-:<5pt>]{\{1,2,5\}} \ar@(ul,u)^(.6){p_2:\{3,4\}}
      \ar[lldd]^(.45){p_1\wedge p_2:\{2,3,4\}} \ar@/_/[rrdd]_(.55){p_2:\{3\}}
      \ar@[red]@/^/[rruu]^(.6){\color{red}p_1\wedge p_2:\{3\}} \\ \\
      &&& *+[F-:<5pt>]{~\{1\}~} \ar@(u,ur)^(.5){\true:\{2,4\}}
      \ar@(dr,d)^(.7){q:\{2,3,4\}} && &&
      *+[F-:<5pt>]{\{1,2,4,5\}} \ar[llll]^{p_1\wedge p_2\wedge p_3:\{2,3,4\}}
      \ar@/_/[lluu]_(.57){p_2\wedge p_3:\{3,4\}} \ar@(dr,d)^(.7){\neg p_1\wedge p_2:\{3\}}
      \ar@<1ex>@/_/[uuuu]_{p_1\wedge p_2\wedge p_3:\{3,4\}} }\]
    \caption{A correct TGBA $\mG'_\varphi$ corresponding to the VWAA of
      Figure~\ref{fig:fin_ex-alt} constructed using modified definition of accepting sets.}
     \label{fig:fin_ex-gen_ok}
  \end{minipage}
\end{figure}


Incorrectness of the improved VWAA to TGBA translation in connection with the original definition of accepting sets is illustrated by the TGBA $\mG_\varphi$ of Figure~\ref{fig:fin_ex-alt} constructed from the VWAA $\mA_\varphi$ of Figure~\ref{fig:fin_ex-gen}. Thanks to the accepting cycle between states $\{1,2,4\}$ and $\{1,2,5\}$, automaton $\mG_\varphi$ accepts the infinite word $w = (\{p_1,p_2\}\{p_3\})^\omega$. Note that $w\not\models\G\F q$ and hence also $w\not\models\psi\wedge\G\F q$. Thus, $w$ is not accepted by VWAA $\mA_\varphi$ as $\mA_\varphi$ corresponds to the formula $\varphi=\psi\wedge\G\F q$. Figure~\ref{fig:fin_ex-gen_ok} depicts the TGBA $\mG'_\varphi$ automaton produced by the improved translation with the new definition of accepting sets. One can easily see that 
$\mG'_\varphi$ does not accept $w$. 

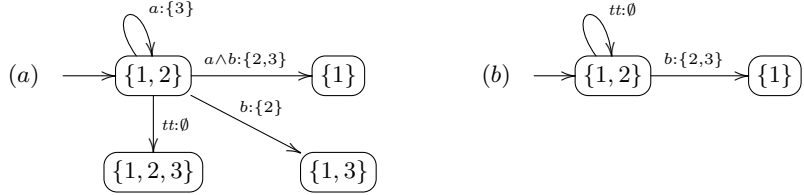
\begin{figure}[tb]
  \[
  \begin{array}{rcrc}
    (a) &
    \xymatrix@C=4ex@R=2ex{
      \ar[r] & *+[F-:<5pt>]{\{1,2\}}
      \ar@(ul,u)^(.7){a:\{3\}} \ar[dd]^{\true:\emptyset} \ar[ddrr]^{b:\{2\}}
      \ar[rr]^{a \wedge b:\{2,3\}} &&
      *+[F-:<5pt>]{\{1\}} \\ \\
      & *+[F-:<5pt>]{\{1,2,3\}}&&
      *+[F-:<5pt>]{\{1,3\}} }
    & ~~~~~~~~~~~~(b) &
    \xymatrix@C=4ex@R=2ex{
      \ar[r] & *+[F-:<5pt>]{\{1,2\}}
      \ar@(ul,u)^(.7){\true:\emptyset} \ar[rr]^{b:\{2,3\}} &&
      *+[F-:<5pt>]{\{1\}} \\ \\
      & *+[]{} }
    \end{array}
    \]
    \caption{Transitions leading from state $\{1,2\}$ in the TGBA
      constructed from the VWAA of Figure~\ref{fig:const_ex-alt} by (a) the
      translation of~\cite{GO01} and by (b) our translation with
      suspension.}
  \label{fig:const_ex}
\end{figure}

To demonstrate the effect of suspension during the construction of a TGBA,
consider the VWAA $\mA_\psi$ for the formula $\psi=\G\F a~\wedge~\F b$
depicted in Figure~\ref{fig:const_ex-alt}. The construction of an equivalent
TGBA $\mG_\psi$ starts in the initial state $\{1,2\}$ that corresponds to a
conjunction of states $1$ and $2$ of $\mA_\psi$. Figure~\ref{fig:const_ex}
depicts the transitions of $\mG_\psi$ leading from the initial state when
constructed by (a) the original translation of~\cite{GO01} and by (b) our
translation with suspension. Note that the state 1 corresponding to the
alternating formula $\G\F a$ is suspended in the TGBA state $\{1,2\}$ as
the state 2 corresponds to a non-alternating progress formula $\F b$. In both
cases, the TGBA has two sets of accepting transitions, $T_2$ and $T_3$. Each
transition in the TGBA is labelled by a propositional formula over $\AP$ and
by a subset of $\{2,3\}$ indicating to which sets of $T_2,T_3$ the
transition belongs.

Comparing to the original VWAA to TGBA translation without any
optimizations, the application of suspension leads to automata with fewer
states. However, if we enable the optimizations suggested in~\cite{GO01},
the original translation often constructs automata with the same number of
states as our translation with suspension. For example, in the TGBA
constructed from the VWAA of Figure~\ref{fig:const_ex-alt}, the
optimizations merge states $\{1,2,3\}$ and $\{1,3\}$ with $\{1,2\}$ and
$\{1\}$, respectively. In this particular case, both approaches lead to the
same automaton $\mG_\psi$ as shown in
Figure~\ref{fig:const_ex-gen_full}. However, this is not the case in
general. Using suspension, automata with either more or less states can be
constructed. However, the translation with suspension is usually slightly
faster.

In addition, we detect that both the original and the improved algorithms
spend a lot of time when computing transitions of TGBA states equivalent to
a formula of the form $\rho = \G\alpha_0~\wedge \bigwedge_{1 \le i \le n}
\G\F\alpha_i$ where $n\ge 0$ and $\alpha_0,\alpha_1,\ldots,\alpha_n$ are
formulae without any temporal operator. As such TGBA states are very
frequent in practice, we use an optimization that detects these TGBA
states and directly constructs the optimal transitions.



\section{Optimization of BA}
\label{sec:tgba2ba}

We slightly modify one optimization rule suggested in~\cite{GO01}. It is
applied on a resulting BA. The rule says that states $q_1$ and $q_2$ of a BA
can be merged if $\delta(q_1)=\delta(q_2)$ and $q_1\in F\Longleftrightarrow
q_2\in F$. This rule typically fails to merge the states with a self
loop. 
We suggest to add a new rule where the condition
$\delta(q_1)=\delta(q_2)$ is replaced by
$\delta(q_1)[q_1/r]=\delta(q_2)[q_2/r]$, where $r$ is a fresh artificial
state and $\delta(q)[q/r]$ is a $\delta(q)$ with all occurrences of $q$ as a
target node replaced by $r$. 


\section{Implementation and Experimental Result}
\label{sec:experiments}

We have implemented all the modifications suggested in the previous sections
(and formula reduction rules suggested in~\cite{EH00}) in order to evaluate
their effect. The implementation is based on LTL2BA and therefore called
\emph{LTL3BA}. Besides the changed algorithms, we also made some other,
implementation related changes. In particular, we represent transition
labels by BDDs and transitions are represented by C++ STL containers.


In this section, we compare LTL3BA with LTL2BA\footnote{Available online at 
\url{http://www.lsv.ens-cachan.fr/~gastin/ltl2ba/index.php}.} (v1.1) and
SPOT\footnote{Available online at \url{http://spot.lip6.fr/wiki/}.}
 (v0.7.1)\footnote{In version version 0.7.1, SPOT
contains a small bug in TGBA degeneralization. We reported this problem to
authors and they provided a corresponding fix which we have
applied. Therefore, the version of SPOT we have actually used differs a bit
from the current version 0.7.1 that is publicly available.}.
For the comparison of results, we use \texttt{lbtt} testbench
tool~\cite{TH02} to measure, for each translator, the number of states and
transitions\footnote{To solve the problem with different representation of
  transitions in automata produced by different tools, we count all
  transitions leading from a state $q$ to a state $r$ as one.} of resulting
automata, and the time of the computation. Further, we extend \texttt{lbtt}
to count the number of produced deterministic automata. To be able to
compare the results, we set SPOT (option \texttt{-N}) to output automata in
the form of never claim for SPIN as that is the output of LTL2BA as
well. All experiments were done on a server with 8 processors
Intel$^\circledR$ Xeon$^\circledR$ X7560, 448~GiB RAM and a 64-bit version
of GNU/Linux.
However, all three translators are single threaded, therefore, they can
utilize only one CPU core.

\begin{table}[t]
  \centering
  \begin{tabular}{|l||c|c|c|c||c|c|c|c|}
\hline
Translator & 
\multicolumn{4}{c||}{Benchmark1} &
\multicolumn{4}{c|}{Benchmark2} \\
\hline
&
States & Trans. & Time & det.~BA & 
States & Trans. & Time & det.~BA \\
\hline \hline
  SPOT
& 1\,561 & 5\,729 & 7.47 & 55
& 14\,697 & 95\,645 & 68.46 & 221 \\
\hline
  SPOT+WDBA
& 1\,587 & 5\,880 & 10.81 & 88
& 13\,097 & 77\,346 & 5\,916.45 & 373 \\
&&&&
& (14\,408) & (94\,248) & (5\,919.43) & (373) \\
\hline
  LTL2BA
& 2\,118 & 9\,000 & 0.81 & 25
& 24\,648 & 232\,400 & 18.57 & 84 \\
\hline
  LTL3BA(1)
& 1\,621 & 5\,865 & 1.26 & 27
& 17\,107 & 129\,774 & 22.25 & 92 \\
\hline
  LTL3BA(1,2)
& 1\,631 & 6\,094 & 1.41 & 54
& 15\,936 & 115\,624 & 9.04 & 237 \\
\hline
  LTL3BA(1,2,3)
& 1\,565 & 5\,615 & 1.41 & 54
& 14\,113 & 91\,159 & 8.53 & 240 \\
\hline
  LTL3BA(1,2,3,4)
& 1\,507 & 5\,348 & 1.38 & 54
& 13\,244 & 85\,511 & 8.30 & 240 \\
\hline
\end{tabular}
 ~\\~
  \caption{Comparison of translators on two sets of random formulae. Time is
    in seconds, 'det.~BA' is the number of deterministic automata produced
    by the translator. Note that, using WDBA minimization,
    SPOT failed to translate 6 formulae of Benchmark2
    within the one hour limit. In order to see the effect of WDBA
    minimization to other formulae, we state in braces the 
    original results increased by the values obtained when these 6 formulae
    were translated withut WDBA minimization.}
  \label{tab:bigs}
\end{table}

First we compare the translators on two sets, Benchmark1 and Benchmark2, of
random formulae generated by \texttt{lbtt}. Benchmark1 contains 100 formulae
of the length 15--20 and their negations. Benchmark2 contains 500 formulae
of the length 15--30 and their negations. The exact \texttt{lbtt} parameters
used to generate the formulae are in Appendix~\ref{subsec:app-lbtt-param}.
Table~\ref{tab:bigs} presents the cumulative results
of translations of all formulae in the two sets. The table also illustrates
the gradual effect of modifications of each step of the translation (1,2,3,4
refers to modifications introduced in
Sections~\ref{sec:reduction},~\ref{sec:ltl2vwaa},~\ref{sec:vwaa2tgba},
and~\ref{sec:tgba2ba} in the respective order; e.g.~LTL3BA(1) uses the
original algorithm with our formula reduction while LTL3BA(1,2,3,4) refers
to the translation with all the suggested modifications). Finally, the table
contains the results for SPOT with WDBA minimization, which has the longest
running time but provides the best results. The automata produced by LTL3BA
are in sum slightly better than the automata produced by SPOT. Further,
LTL3BA seems to be much faster.



\begin{figure}[tb]
	\includegraphics[scale=0.8]{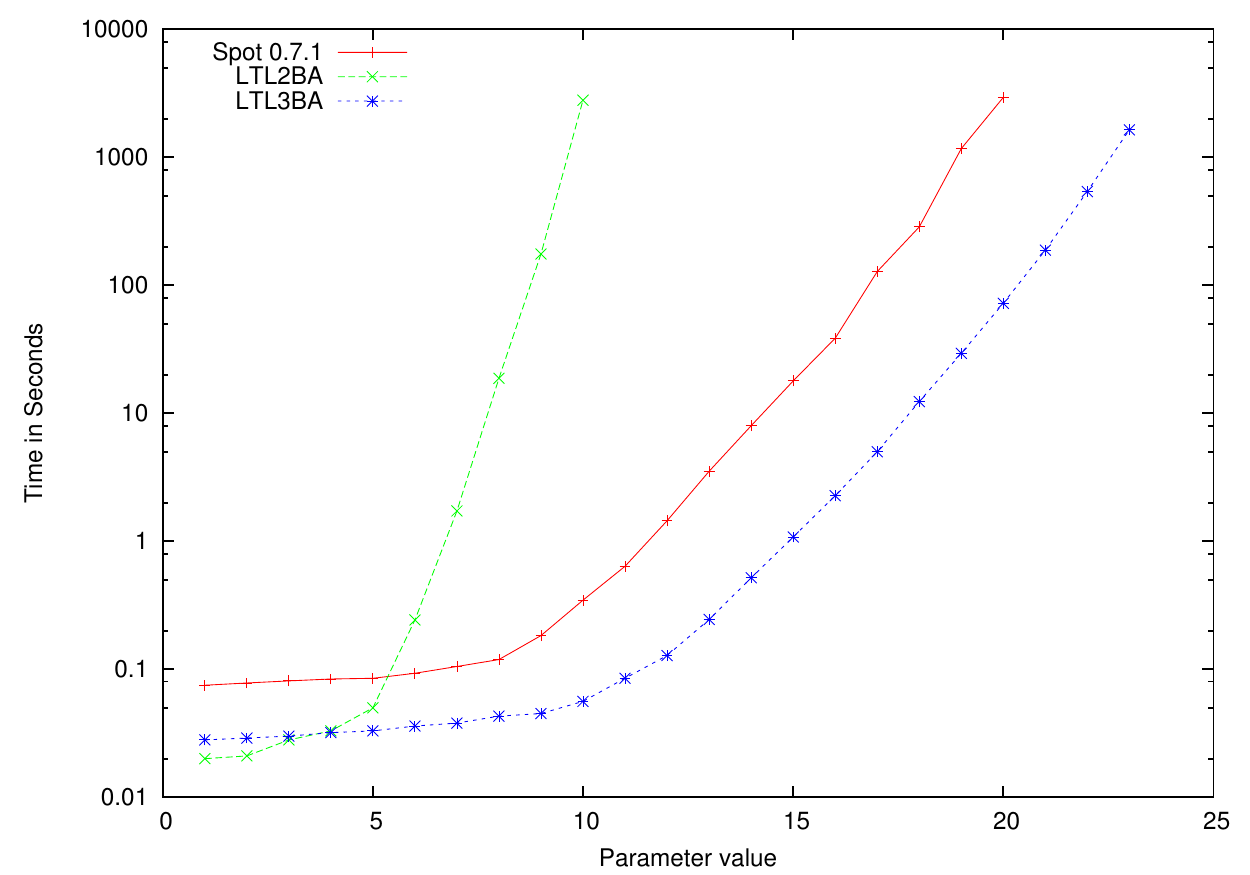}
    \\ \centering $\theta_n = \neg((\G\F p_1 \wedge \ldots
      \wedge \G\F p_n) \rightarrow \G(q \rightarrow \F r))$
  \caption{Running times of LTL to BA translators on parametric formula $\theta_n$
  	of~\cite{GO01} (the vertical axe is logarithmic and represent time in seconds,
    while the horizontal axe is linear and represent the
    parameter $n$).}
  \label{fig:graf_theta}
\end{figure}

\begin{figure}[tb]
   \begin{minipage}[b]{0.495\linewidth}
    \includegraphics[width=\linewidth]{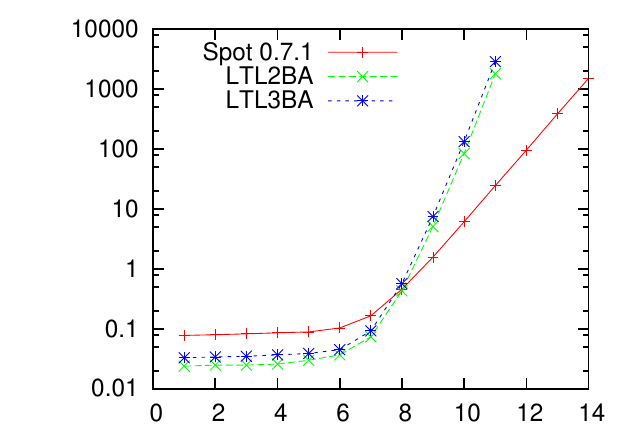}
    \\ \centering $~~~~~~~~~U_1(n) = (\ldots(p_1 \U p_2) \U \ldots) \U p_n$
  \end{minipage}
  \begin{minipage}[b]{0.495\linewidth}
    \includegraphics[width=\linewidth]{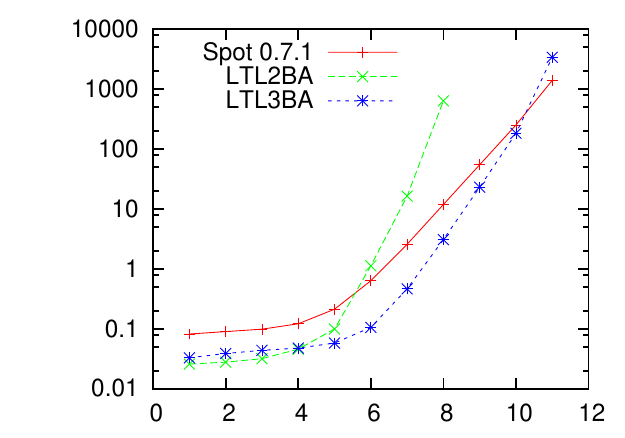}
    \\ \centering $~~~~~~~~~R(n) = \bigwedge^n_{i=1} (\G\F p_i \vee
      \F\G p_{i+1})$
  \end{minipage}
  \\[2mm]
  \begin{minipage}[b]{0.495\linewidth}
    \includegraphics[width=\linewidth]{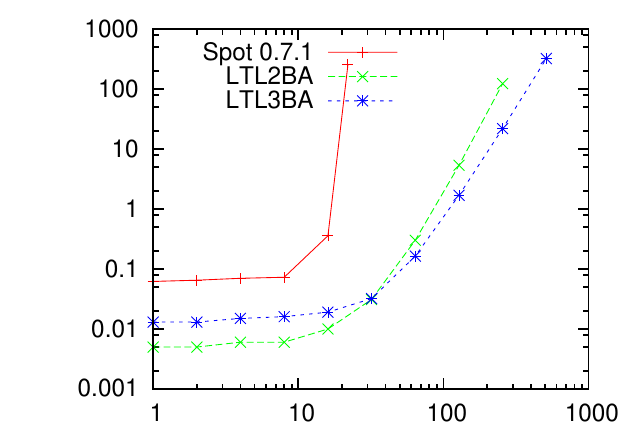}
    \\ \centering $~~~~~~~~~U_2(n) = p_1 \U (p_2 \U (\ldots p_{n-1} \U p_n) \ldots)$
  \end{minipage}
  \begin{minipage}[b]{0.495\linewidth}
    \includegraphics[width=\linewidth]{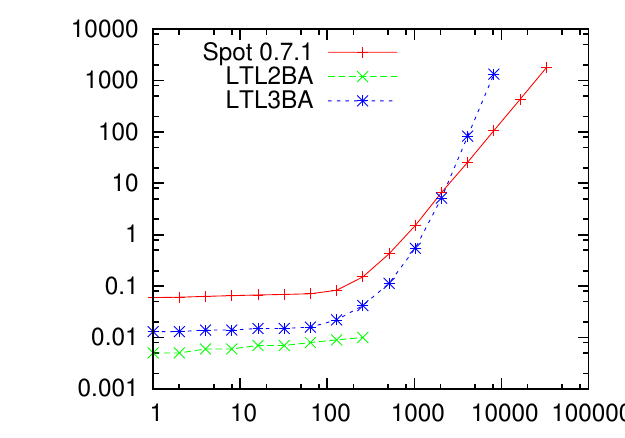}
    \\ \centering $~~~~~~~~~S(n) = \bigwedge^n_{i=1} \G p_i$
  \end{minipage}
  \caption{Running times of LTL to BA translators on parametric formulae
  	of~\cite{RV07} (the vertical axes are logarithmic and represent time in seconds,
    while the horizontal axes are linear or logarithmic and represent the
    parameter $n$).}
  \label{fig:grafy}
\end{figure}

\begin{figure}[tbh]
  \begin{minipage}[b]{0.495\linewidth}
    \includegraphics[width=\linewidth]{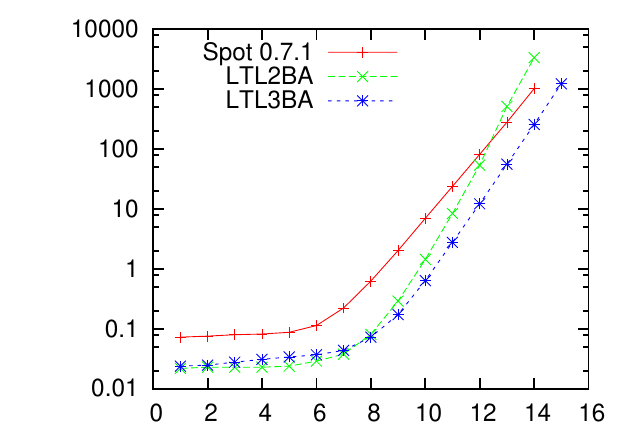}
    \\  \centering $~~~~~~~~~E(n) = \bigwedge^n_{i=1} \F p_i$ 
  \end{minipage}
  \begin{minipage}[b]{0.495\linewidth}
    \includegraphics[width=\linewidth]{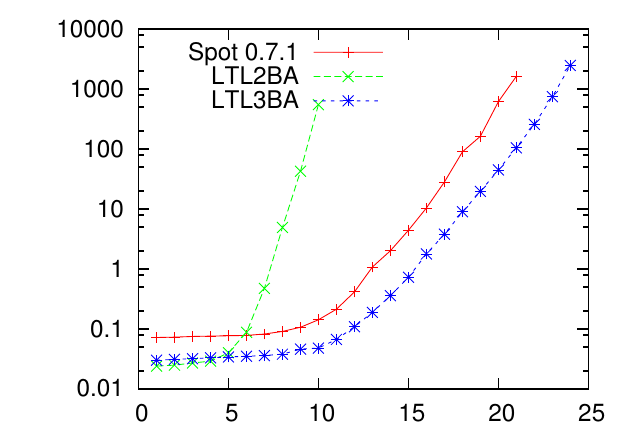}
    \\ \centering $~~~~~~~~~C_2(n) = \bigwedge^n_{i=1} \G\F p_i$
  \end{minipage}
  \\[2mm]
  \begin{minipage}[b]{0.495\linewidth}
    \includegraphics[width=\linewidth]{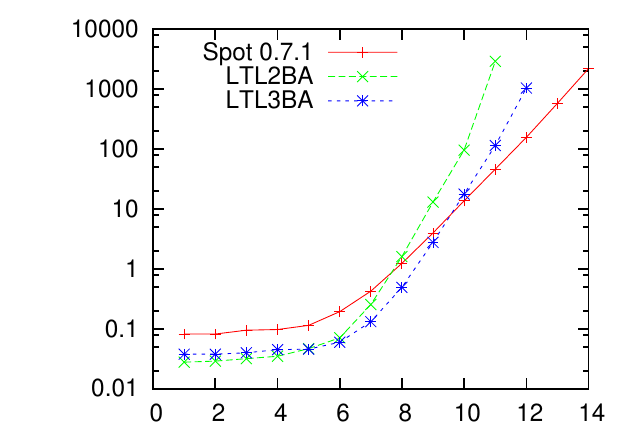}
    \\ \centering $~~~~~~~~~Q(n) = \bigwedge^n_{i=1} (\F p_i \vee \G
      p_{i+1})$
  \end{minipage}
  \begin{minipage}[b]{0.495\linewidth}
    \includegraphics[width=\linewidth]{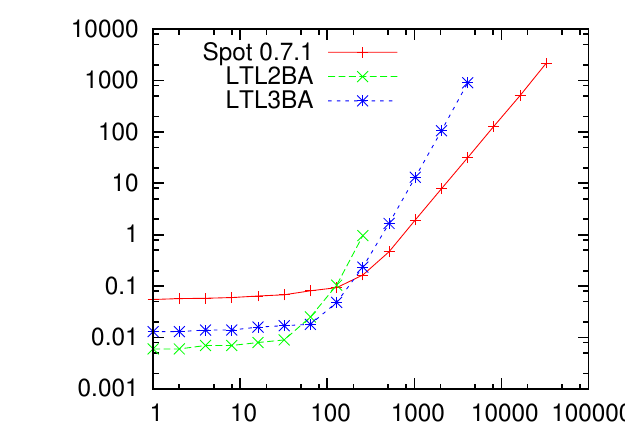}
    \\ \centering $~~~~~~~~~C_1(n) = \bigvee^n_{i=1} \G\F p_i$
  \end{minipage}
  \caption{Running times of LTL to BA translators on another parametric formulae
  	of~\cite{RV07} (the vertical axes are logarithmic and represent time in seconds,
    while the horizontal axes are linear or logarithmic and represent the
    parameter $n$).}
  \label{fig:grafy_app}
\end{figure}

Further, we compare the execution time of translators running on parametric
formulae from~\cite{GO01} and~\cite{RV07}. We use SPOT with the recommended
option \texttt{-r4}, i.e.~with the input formula reduction as the only
optimization. To get a comparable settings of LTL3BA, we switched off the
generalized optimization of VWAA. We gradually increase the parameter of the
formulae until a translator fails to finish the translation in one hour
limit. The results are depicted in Figure~\ref{fig:graf_theta}, Figure~\ref{fig:grafy} and Figure~\ref{fig:grafy_app}.

It is worth mentioning that each automaton produced by LTL3BA for $\theta_n$
has around half the number of states and half the number of transitions than
the one produced by SPOT. If we use default settings for SPOT, the automata
are of the same size as from LTL3BA and the maximal formula is $\theta_{20}$
too. The other way round, if we add SCC based automata simplification
(option \texttt{-R3}) the results are small again but the maximal formula
computable in an hour is $\theta_{18}$.

The graphs show that, in general, LTL3BA is slightly slower than LTL2BA and
faster than SPOT on small formulae. With increasing parameter, LTL3BA
outperforms LTL2BA (with exception of $S(n)$ where LTL2BA fails before its
running time reaches the limit), while SPOT sometimes remains slower, but
sometimes eventually outperform LTL3BA.

Finally, we compared SPOT and LTL3BA on parametric formulae from
\cite{Cichon_2009}:
\[
\begin{array}{rcl}
  \alpha_n &=& \F (p_1 \wedge \F(p_2 \wedge \ldots \wedge \F p_n) \ldots) \wedge \F (q_1 \wedge \F(q_2 \wedge \ldots \wedge \F q_n) \ldots) \\
  \beta_n &=& \F (p \wedge \X(p \wedge \ldots \wedge \X p) \ldots) \wedge \F (q \wedge \X(q \wedge \ldots \wedge \X q) \ldots) \\
  \beta'_n &=& \F (p \wedge \X p \wedge \X^2 p \wedge \ldots \wedge \X^{n-1} p) \wedge \F (q \wedge \X q \wedge \X^2 q \wedge  \ldots \wedge \X^{n-1} q) \\
  \psi_n &=& \G\F p_1 \wedge \G\F p_2 \wedge \ldots \wedge \G\F p_n \\
  \xi_n &=& \F\G p_1 \vee \F\G p_2 \vee \ldots \vee \F\G p_n \\
\end{array}
\]
In 2009, Cicho{\'n} et al.~\cite{Cichon_2009} introduced the four parametric
formulae and shown that their BA representations obtained by both LTL2BA and
SPOT are far away from their minimal representations (or uncomputable even
for the parameter $n \leq 20$). Two years later in~\cite{Lut11}, the authors
of SPOT announced that they are able to compute all the mentioned formulae
in minimal form. We have recomputed the results for all $n\leq 20$ by SPOT
and LTL3BA and realized that LTL3BA returns also the minimal automata but
8 times faster. More precisely, the overall computation of SPOT took more than 13 minutes (802 seconds), while the computation of LTL3BA took less than 2
minutes (95 seconds).

\section{Conclusion}\label{sec:concl}

We have focused on LTL to BA translations with the stress on their speed-up
while maintaining outputs of a good quality. We have introduced several
modifications of LTL2BA on both algorithmic and implementation levels.
Among others, we have identified an LTL subclass of ``alternating''
formulae, validity of which does not depends on any finite prefix of the
word.

Our experimental results indicate that our modifications have a mostly
positive effect on each step of the translation. The new translator called
LTL3BA is usually faster than the original LTL2BA and it produces smaller
and more deterministic automata. Moreover, comparison of LTL3BA and the
current version of SPOT (run without WDBA minimization that is very slow)
shows that the produced automata are of similar quality and LTL3BA is
usually faster.

LTL3BA has served as an experimental tool to demonstrate that our modifications
are improvements and their applicability to other LTL to BA translations is
a subject of further research.

LTL3BA is publicly available under GPL at:
\begin{center}
\fbox{~\url{http://sourceforge.net/projects/ltl3ba/}\strut~}
\end{center}


\paragraph{Acknowledgments.} The authors would like to thank three anonymous
refrees and Alexandre Duret-Lutz for valuable comments. 


\bibliographystyle{plain}

\newpage
\appendix


\section{\texttt{lbtt} parametters used for formulae generation}\label{sec:app}

%
%

\label{subsec:app-lbtt-param}
Here are the precise parameters for \texttt{lbtt} to produce the sets 
Benchmark1 and Benchmark2 of random formulae. Note that we also added
negations of these formulae to the sets. 

Benchmark1 (100 formulae + their negations):
\begin{verbatim}
  Size = 15...20
  Propositions = 8

  AbbreviatedOperators = Yes
  GenerateMode = Normal
  OutputMode = NNF
  PropositionPriority = 50

  TruePriority = 1
  FalsePriority = 1

  AndPriority = 10
  OrPriority = 10
  XorPriority = 0
  EquivalencePriority = 0

  BeforePriority = 0
  StrongReleasePriority = 0
  WeakUntilPriority = 0

  UntilPriority = 30
  DefaultOperatorPriority = 15
\end{verbatim}
The parameters for Benchmark2 (500 formulae + their negations) are the same
except the first one, where set:
\begin{verbatim}
  Size = 15...30
\end{verbatim}

\end{document}